\documentclass[twocolumn,showpacs,preprintnumbers,amsmath,amssymb, superscriptaddress,prl]{revtex4}
\pdfoutput=1
\usepackage{latexsym}
\usepackage{amssymb}
\usepackage{epsfig,amsmath,graphics}
\usepackage{epstopdf}

\begin{document}

\preprint{SLAC-PUB-14558, SU-ITP-11/46}

\title{A Simple Harmonic Universe}

\author{Peter W. Graham}
\affiliation{Stanford Institute for Theoretical Physics, Department of Physics, Stanford University, Stanford, CA 94305}

\author{Bart Horn}
\affiliation{Stanford Institute for Theoretical Physics, Department of Physics, Stanford University, Stanford, CA 94305}
\affiliation{SLAC National Accelerator Laboratory, Stanford University, Menlo Park, California 94025}

\author{Shamit Kachru}
\affiliation{Stanford Institute for Theoretical Physics, Department of Physics, Stanford University, Stanford, CA 94305}
\affiliation{SLAC National Accelerator Laboratory, Stanford University, Menlo Park, California 94025}

\author{Surjeet Rajendran}
\affiliation{Department of Physics and Astronomy, The Johns Hopkins University, Baltimore, MD, 21218}
\affiliation{Stanford Institute for Theoretical Physics, Department of Physics, Stanford University, Stanford, CA 94305}

\author{Gonzalo Torroba}
\affiliation{Stanford Institute for Theoretical Physics, Department of Physics, Stanford University, Stanford, CA 94305}
\affiliation{SLAC National Accelerator Laboratory, Stanford University, Menlo Park, California 94025}

\date{\today}

\begin{abstract}
We explore simple but novel bouncing solutions of general relativity that avoid singularities.  
These solutions require curvature $k=+1$, and
are supported by a negative cosmological term and matter with $-1 < w < -1/3$.
In the case of moderate
bounces (where the ratio of the maximal scale factor $a_+$ to the minimal scale factor $a_-$ is ${\cal O}(1)$), 
the solutions are shown to be classically stable and cycle through an infinite set of bounces.
For more extreme cases with large $a_+/a_-$, the solutions can still oscillate many times before classical instabilities take them out of the regime of validity of our approximations.
In this regime, quantum
particle production also leads eventually to a departure from the realm of validity of semiclassical general relativity,
likely yielding a singular crunch. We briefly discuss possible applications of these models to realistic cosmology.

\end{abstract}

\maketitle

Two questions have recurred often in theoretical cosmology \cite{previous works}:\ 1) is the Universe eternal or
did it have a beginning at some definite time in the past?, and 2) is it possible to make
Universes with one or more ``bounces" where the scale factor crunches
and then bangs? \footnote{We discuss solutions where the universe is parametrically larger than the Planck length.
Ambitious models with crunches require a boundary condition at the singularity provided by the unknown high energy theory \cite{PBB,EK}. More recent works on related ideas that also analyze perturbations around the bounce include
\cite{extraNECviol}.}.

The answers to these two questions
are deeply intertwined with the subject matter of the singularity theorems of Penrose and Hawking
(discussed comprehensively in \cite{HawkingEllis}).
These theorems show that, given an energy condition of the form
\begin{equation}
\label{eqn:EC}
T_{\mu\nu}v^{\mu}v^{\nu} \geq \frac{1}{2} T v_\mu v^\mu
\end{equation}
for a suitable class of vectors $v^\mu$, where $T_{\mu\nu}$ is the
stress-energy tensor of the sources supporting the Universe,
one can prove that the Universe must be 
geodesically incomplete (``singular").  Even in scenarios where the current $\Lambda {\rm CDM}$ cosmology was preceeded by a phase of slow-roll inflation \cite{Inflation}, with eternal inflation occurring on even larger scales, it is striking \cite{Guth} that the initial singularity remains, independent of the energy condition assumed.

It is instructive to discuss which energy conditions need to be assumed to 
prove existence of a cosmological singularity for the FLRW cosmologies
\begin{equation}
\label{eqn:FRW metric}
ds^2 = -dt^2 + a(t)^2 \left( \frac{dr^2}{1-kr^2} + r^2 (d\theta^2 + {\rm sin}^2(\theta) d\phi^2) \right).
\end{equation}
For $k=-1, 0$ the only condition that must be assumed is the null energy condition (NEC), i.e.~eqn.~\eqref{eqn:EC} where $v^\mu$ is a
future-pointing null vector field.
The NEC is reasonable and in agreement with the known macroscopic matter and energy
sources in our Universe \footnote{Interesting cosmological scenarios which attain a smooth bounce by
violating the NEC can be found in \cite{Senatore}.}.

For $k=+1$, however, the ${\it strong}$ energy condition (SEC) (where $v^\mu$ in (\ref{eqn:EC}) is future-pointing timelike) must be assumed \footnote{More generally, the power of the singularity theorems is that they apply
for much less symmetric space-times than those allowed by the FLRW ansatz.  In these generic cases as well, the SEC must be assumed to prove a theorem.  So our  results for $k=+1$ FLRW may be reflective of phenomena that can occur in less symmetric space-times.}.
This condition is violated by macroscopic sources
in our world, as well as in many completely consistent theoretical toy models.  Our goal is to explore the two questions above for $k=+1$ Universes with 
sources satisfying the NEC (but violating the SEC).  We will find that one can make
classical cosmologies that  live
eternally, undergoing an infinite sequence of non-singular bounces, and remaining within the regime of validity 
of general relativity. When the ratio between maximal and minimal scale factors is not too large, these cosmologies are stable to small perturbations.
When
the ratio is large, we instead find both classical and
quantum pathologies; classically there are growing modes (which can be tuned away), and quantum mechanically,
particle production backreacts significantly, likely causing a singular crunch.

{\it Solutions.} 
The FRW equations for the metric eqn.~\eqref{eqn:FRW metric} are
\begin{equation}
\label{eqn:FRW1}
\frac{\dot{a}^2}{a^2} = \frac{8 \pi}{3} G \rho - \frac{k}{a^2}\;\;,\;\;\frac{\ddot{a}}{a} = -\frac{4 \pi}{3} G \left( \rho + 3 p \right)
\end{equation}
where $\rho$ is the energy density and $p$ is the pressure.
We want oscillatory solutions, namely those with two extrema ($\dot{a} = 0$) such that at the smaller (where $a \equiv a_{-}$) $\ddot{a} > 0$, and at the larger (where $a \equiv a_{+}$) $\ddot{a} < 0$.  It is easy to see that these requirements, along with the NEC, only allow solutions for $a$ when there is positive curvature, $k = +1$.  The minimal model which oscillates has three components:\ positive curvature, a negative cosmological constant (energy density $= \Lambda < 0$), and a ``matter" source with equation of state in the range
\begin{equation}
\label{weneed}
p = w\rho, ~-1 < w < -1/3
\end{equation}
(we will see later that this must ${\it not}$ be a perfect fluid).  For this content the energy density is $\rho = \Lambda + \rho_0 \, a^{-3 (1+w)}$ where $\rho_0$ is a constant parametrizing the density of the ``matter."  Then the solution to eqns.~\eqref{eqn:FRW1} is oscillatory.

In the special case that $w=-\frac{2}{3}$ these equations just describe a constrained simple harmonic oscillator and the solution (setting $k=+1$) is \footnote{Note added: Shortly after the first version of this pa- per appeared, we learned of the related works \cite{previous} where the same solution is described at the homogeneous level. These authors did not discuss the stability questions raised by multiple bounces.
}
\begin{equation}
\label{scalesol}
a = \frac{\rho_0}{2 | \Lambda |} + a_0 \cos \left(\omega t  + \psi \right)
\end{equation}
where $\psi$ is an arbitrary phase and
\begin{equation}
\omega \equiv\sqrt{ \frac{8 \pi}{3} G |\Lambda|}\;,\; a_0 \equiv \frac{1}{2 |\Lambda|} \sqrt{\frac{3 \Lambda}{2 \pi G} + \rho_0^2}.
\end{equation}
This requires $\rho_0^2 \ge \frac{3}{2 \pi} \frac{|\Lambda|}{G}$ for positivity of the radicand.  Note that the Universe is static when this condition is saturated, though this requires a fine-tuning.  In the opposite limit, $\frac{\rho_0^2}{\Lambda} \to \infty$, the ratio of the maximum to the minimum sizes $a_+/a_-$ of the Universe goes to infinity.

It is useful to switch to conformal time $\eta$, where $d\eta^2 =dt^2 /a(t)^2$. Defining
\begin{equation}
\label{gammais}
\gamma \equiv \frac{3|\Lambda|}{2\pi G \rho_0^2}
\end{equation}
the solution for the scale factor (\ref{scalesol}) becomes
\begin{equation}
\label{confscale}
a(\eta) = {1\over \omega} {\sqrt{\gamma} \over {1 - \sqrt{1-\gamma} \, {\rm cos}(\eta)}}~.
\end{equation}
Here $\omega$ is the frequency of oscillations given in (\ref{scalesol}), and we have set $\psi=0$. Notice that $\gamma \approx 4 a_-/a_+$ for small $\gamma$.

{\it Stability.} There are several simple stability issues we discuss here.  (See e.g.\ \cite{Barrow:2003ni} for a discussion of the corresponding stability issues in the Einstein static Universe.)  First of all, the ``matter" source in Eqn.~\eqref{weneed}
may itself present dangers.  In fact the canonical source which behaves this way, a perfect fluid, would present a serious problem.  To see this, recall that for scalar perturbations, one considers a more general metric
\begin{equation}
ds^2 = a(\eta)^2\left[- (1+ 2\Phi(\eta,x)) d\eta^2 +  (1- 2\Psi(\eta,x)) d\Omega_3^2\right].
\end{equation}
For perfect fluids, $\Phi = \Psi$, $\delta p = c_s^2 \delta \rho$, and
\begin{eqnarray}
\label{Psidifeq}
 \Psi'' + 3\mathcal{H} (1+c_s^2) \Psi' &+& \left[ 2 \mathcal{H}' + (1+3c_s^2) (\mathcal{H}^2-k)\right]\Psi\nonumber\\
 & -&c_s^2 \nabla^2_{S^3} \Psi = 0~.
\end{eqnarray}
The derivatives are with respect to conformal time, and $\mathcal{H}=a'/a$.
As is clear from the ${\it sign}$ of the $ \nabla^2_{S^3}$ term in \eqref{Psidifeq}, if $c_s^2 < 0$, high-momentum modes are unstable.

Now, a perfect fluid with $w < -1/3$ would have negative $c_s^2.$  However, as explained in \cite{Spergel}, one can find matter sources supporting equations of state of the form (\ref{weneed}) but with $c_s^2 > 0$ (and in fact comparable to the speed
of light), if one considers a ``solid" with elastic resistance to shear deformations.  A canonical example which they discuss is a frustrated network of
domain walls, which in the leading approximations gives precisely the simple $w=-{2\over 3}$ case.
For our purposes, the crucial point is simply that once we have achieved
$c_s^2$ sufficiently positive, it is easy to check that the scalar perturbations above are stable.

In addition to the above scalar perturbations, we need to consider tensor perturbations. These are governed by an equation whose form is identical to that of \eqref{scaleeom} below, and will be analyzed there. Next, homogenous but anisotropic perturbations are given by the Bianchi type IX metric \cite{bianchi} $ds^2 = -dt^2 + \sum_{i=1}^3\,a_i^2(t)\,\sigma_i^2$, where $\sigma_i$ are the Maurer-Cartan forms on $S^3$. It is useful to parametrize the $a_i$ by an overall $a(t)$ and two `shape' deformations $\beta_{\pm}(t)$,
\begin{equation}
a_1 = a\, e^{\frac{\beta_+ + \beta_-}{2}}\;,\;a_2 = a\, e^{\frac{\beta_+ - \beta_-}{2}}\;,\;a_3= a\, e^{-\beta_+}\,.
\end{equation}
Linearizing the FRW equations for $\beta_{\pm} \ll 1$ then obtains
\begin{equation}\label{mixmaster}
\beta_{\pm}''+ 2 \mathcal{H} \beta_{\pm}'+ 8 k\beta_{\pm}=0\,.
\end{equation}
These modes will be analyzed momentarily.

Another potential source present in our Universe is gravity itself, e.g.~a produced
gas of gravitons.  The dynamics of massless particles may be described by a probe scalar field, with equation of motion
\begin{equation}
\label{scaleeom}
\phi^{\prime\prime} + 2 \mathcal{H} \phi^\prime -\nabla^2_{S^3} \phi ~=~0\,.
\end{equation}
Interestingly, because of the periodicity of $a$, (\ref{Psidifeq}) and (\ref{scaleeom}) can be recast as a Schr\"odinger problem in a particular 1d periodic potential.

The three types of perturbations \eqref{Psidifeq}, \eqref{mixmaster} and \eqref{scaleeom} have a similar structure; in fact, the anisotropic perturbation \eqref{mixmaster} is just a particular case of \eqref{scaleeom}. Tensor modes of the metric are also described by eqn.~\eqref{scaleeom}. We denote a generic linearized mode by $u$, and expand in spherical harmonics, $ \nabla^2_{S^3} u_l=-l(l+2) u_l$. We now summarize the results of our numerical analysis of perturbations.

There are
three regimes of momenta where we will find different
behaviors.  It is important to distinguish Universes with $\gamma \sim {\cal O}(1)$ from
those with $\gamma \ll 1$; we describe the behavior in both limits.

\noindent$\bullet$ $l=0$ homogeneous mode: shifting such a mode should be analogous to shifting the homogeneous mode of the scale factor, which would simply
move us in the space of periodic solutions and lead to a linear growth of the perturbation in naive perturbation theory (since e.g.\ two sinusoidal functions
with slightly different frequency will perturbatively grow apart at a linear rate, as they get out of phase).  This is borne out by the numerics for both
$\gamma \ll 1$ and $\gamma \sim 1$.  So what looks like a growing perturbation is likely just a failure of perturbation theory.
  
\noindent$\bullet$ modes with momentum $2\le l \lesssim {1\over \sqrt{\gamma}}$ on the $S^3$:  these are long enough to detect the
difference between our cosmology and Minkowski space. 
For $\gamma \sim 1$, i.e.\ a Universe ``quivering" around a mean size, they are oscillatory and stable.
 In contrast, for $\gamma \ll 1$, they can be unstable; we shall discuss bounds derived from their behavior
below.

\noindent$\bullet$ modes with $l \gg  {1\over \sqrt{\gamma}}$: these have small enough wavelength that they
should barely detect the departures of our metric from flat space. As expected, they behave like typical Minkowski space scalar field modes for times smaller than the period of oscillation of the Universe, for both $\gamma \ll 1$ and $\gamma \sim 1$.

The $l=1$ mode is special. The perturbations governed by \eqref{scaleeom} are stable for $\gamma \sim 1$; on the other hand, the gravitational instabilities sourced by \eqref{Psidifeq} are always unstable for $l=1$.  For the case of a single-component perfect fluid, on which we have focused so far, this mode is absent from the physical spectrum: $\partial_i \Psi_{l=1}$ generates a global rotation on the $S^3$ and hence is pure gauge. However, in multi-component systems there will generically be entropy perturbations; these contribute an inhomogeneous term to (\ref{Psidifeq}) and can source a physical $l=1$ mode.  We find that the corresponding metric scalar mode $\Psi_{l=1}$ grows for all $\gamma$, unlike the case of modes with $l \ge 2$. \footnote{The metric scalar perturbations are classified as adiabatic (curvature) or entropy (isocurvature). The later are described by (\ref{Psidifeq}) plus an inhomogeneous term, and satisfy the initial condition $\Psi= \Psi'=0$. This term can act as a source of $\Psi_{l=1}$, which grows exponentially for all $\gamma$.} 

However, even in these cases the $l=1$ mode may be absent due to different mechanisms. A simple variant of our setup would be to orbifold the $S^3$ by a freely acting group in order to project out this mode. Orbifolding does not change the equations of motion but will project out modes from the spectrum. Further, non-gravitational damping must be included. Collisionless damping (free streaming) occurs at a rate proportional to the frequency $\omega_k$ of the mode. There is a range of $\gamma$ for which the $l=1$ mode predicted by \eqref{Psidifeq} is completely killed by free streaming. The other fluids in the setup, including the domain wall network, may also have other collisional forms of damping that can reduce the growth of this mode.
In what follows we will assume that the $l=1$ growing mode is absent.

\begin{figure}
\begin{center}
\includegraphics[width=2in]{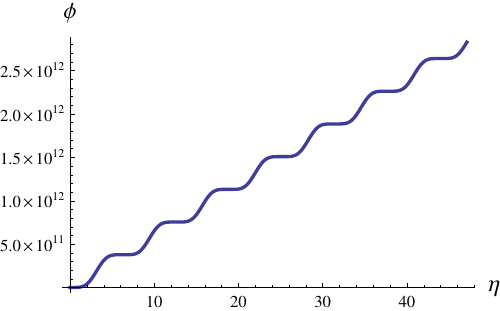}
\includegraphics[width=2in]{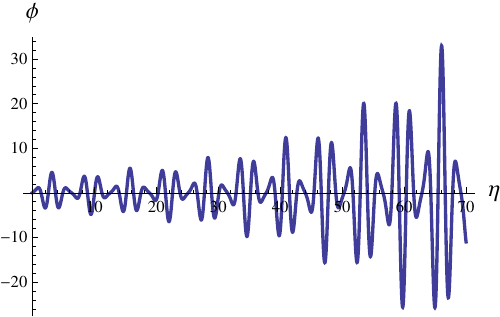}
\includegraphics[width=2in]{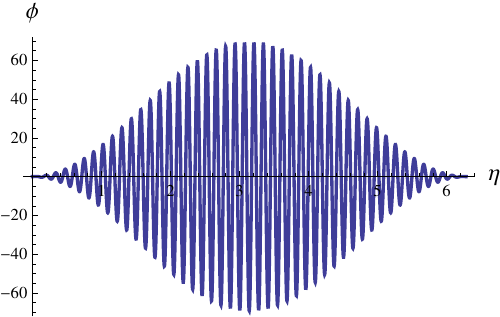}
\caption{ \label{fig:pert} Massless scalar field evolution in conformal time, for different values of momenta. The first plot shows the homogeneous ($l=0$) solution with $\gamma=10^{-5}$. The second plot corresponds to $l=2$ and $\gamma=0.225$; three cycles are included, showing the exponential growth in the amplitude. The third plot has $l=45$ and $\gamma=0.01$, and shows a single cycle. The initial conditions are $\phi(0)=0$ and $\phi'(0)=1$. The exponential growth whose beginning is shown in the middle figure would not be present for $\gamma \sim 1$.}
\end{center}
\end{figure}

To summarize, the Universes with $\gamma \sim 1$ are classically stable at the linearized level and live forever.  The Universes with $\gamma \ll 1$ suffer from exponential growth (as a function of cycle number)
of the finite momentum modes with $l \ll {1\over \sqrt{\gamma}}$.
We show the numerical analysis of the modes of eqn.~\eqref{scaleeom} in Figure \ref{fig:pert} for all three regimes of momenta and various values of $\gamma$. The metric scalar perturbations $\Psi$ behave in a qualitatively similar way, although they exhibit a faster growth rate due to the gravitational backreaction included in eqn.~\eqref{Psidifeq} \footnote{As a check we note that the homogeneous equation can be solved exactly, exhibiting the expected linear growth.  The other behaviors are similarly as one expects, and the crossover between the linearly growing, exponentially growing, and well-behaved short-wavelength modes occurs smoothly, giving no indication of numerical glitches.}.

{\it Classical and quantum destruction of the Universe.} 
For $\gamma \sim 1$, the Universes we are studying are classically stable.  For $\gamma \ll 1$,
the exponential growth of the modes with $0 < l < {1\over \sqrt{\gamma}}$ clearly indicates that we should expect such a Universe to have a bounded lifetime. Can we tune this to allow a large number of oscillations within our regime of computational control?

The cross-over from exponential to oscillatory behavior in the numerical solutions at $l \sim l_c = {1\over \sqrt{\gamma}}$, together with basic attempts to fit the 
growing solutions, suggest a rough form for the growing modes
\begin{equation}
\label{phigrowth}
u_{l}(N) \sim u_0 ~{\rm exp}\left( c \sqrt{1 - {l^2 \over l_c^2}} \times N\right)
\end{equation}
where $c \sim O(1)$, and $u_l(N)$ denotes the value of the $l$th momentum mode after $N$ oscillations, with starting vev $u_0$.
We will compute when these modes grow sufficiently to dominate the energy density, thus altering our solution.
The ratio at $a_-$
of the energy density in the scalar perturbation to the domain wall network is given by
\begin{equation}
\sum_l {{a^2 l (l+2) u_l^2} \over {a^3 \rho_0}} \sim {1 \over M_P^2} \int^{l_c} ~dl~l^2 u_l^2~.
\end{equation}
Using (\ref{phigrowth}), and evaluating the resulting integral in a saddle-point approximation, we find the dominant $l$ is $l_{saddle}^2 \sim l_c^2/N$,
and the energy ratio is thus
\begin{equation}
{\epsilon_u \over \rho} \sim l_c^2 {u_0 ^2 \over M_P^2}~ {\rm exp}\left( O(N) - O({\rm log} N)\right)~.
\end{equation}
So, backreaction from the classical scalar field becomes important after a number of cycles $N_c$ given by
\begin{equation}
N_c \sim ~{\rm log} \left( {M_P^2 \gamma^{3/2}} \over{ u_0^2} \right)~.
\end{equation}
Classically, by tuning $u_0$ to be small, we can obtain an arbitrarily large lifetime even for the
systems with $\gamma \ll 1$.

Quantum mechanics is expected to induce an RMS value of $u_0$, preventing a classical tune from saving the Universe for $\gamma \ll 1$.  Consider the scalar field (\ref{scaleeom}).
To quantize the field, we impose canonical commutation
relations on the canonically normalized scalar $\chi \equiv a(\eta) \phi$, 
\begin{equation}
[ \chi(\theta), \partial_{\eta} \chi (\theta^\prime) ] =~i\delta^{(3)}(\theta - \theta^\prime)\,,
\end{equation}
where $\theta$ coordinatizes the three-sphere.  This implies that in the instantaneous ground state characterizing the scalar at a time when the Universe has scale factor $a$, $a^2 \phi_0^2 \sim 1$.  Now $a_+ = {2\over \omega \sqrt{\gamma}} $ and $a_- = {\sqrt{\gamma} \over{2 \omega}}$.   We may choose, as our initial quantum state,
the instantaneous vacuum associated to any value of the scale factor.  Choosing, for instance, the ``natural"
quantum vacuum associated with $a=a_+$ (where the Universe is large and smooth and we have a natural expectation for the vacuum
state), gives  $\phi_0 \sim \omega \sqrt{\gamma}$. This gives a bound on the number of cycles
\begin{equation}
N_c \sim {\rm log} \left( {M_P  \sqrt{\gamma}} \over {\omega^2} \right)~.
\end{equation}
This can be made parametrically large for small values of $\Lambda$.

For $\gamma \sim 1$ the solutions to \eqref{scaleeom} are oscillatory, so the RMS values for various fields induced by quantum mechanics will not cause instabilities. Hence for these values of $\gamma$, the universe is stable against perturbative classical and quantum instabilities. 

{\it Non-linear instabilities.} The above analysis  of stability has been performed at the linear level. One may ask if this stability would persist at the non-linear level. Non-linear interactions could cause the oscillating scale factor to excite higher energy modes of the system. Such excitations will lead to the continuous production of entropy, destroying the periodic nature of the solution, potentially leading to a crunch initiated by the produced entropy. In the context of the present work, this discussion is pertinent in the case $\gamma \sim 1$ where we have stability at the linear level. While there are non-linear couplings between the oscillating scale factor and higher energy excitations, these couplings result in excitations only when the higher energy excitations are sufficiently close to an integer multiple of the frequency of the oscillating scale factor \cite{Matthieu}. A priori, the higher energy excitations in this system are not integer multiples of the oscillating scale factor. Hence it is possible that the system could be completely stable at the non-linear level. It is also possible, however, that a linear combination of these higher energy excitations may be sufficiently close to an integer multiple of the frequency, resulting in continuous excitation of such modes and entropy production.

In general this is a difficult question, as evidenced by the fact that the non-linear stability of Minkowski space was only recently established \cite{Christodoulou:1993uv}. Such an analysis is thus beyond the scope of this work. We point out though that the case of the oscillating universe with $\gamma \sim 1$ is quite different from typical thermodynamic systems where we expect continuous entropy production. In a typical thermodynamic system, there are a large number of modes that are roughly degenerate with some initial excitation. Due to this large degeneracy, it is relatively easy to satisfy the conditions necessary for efficient excitation of other modes through non-linear interactions. In the case of the oscillating universe with $\gamma \sim 1$, however, the modes that can be excited by such non-linear couplings are at higher energy. Furthermore, as the degeneracy of the modes increases, so does their energy, in contrast with typical thermodynamic systems where there are a large number of low energy modes.  We are unaware of concrete arguments that establish the generation of entropy in systems that share the spectrum of this oscillating universe. One might expect such high energy modes to decouple from the low frequency excitations of the scale factor, potentially leading to an eternal universe.

{\it Nonperturbative instabilities.} Another class of instabilities arise from nonperturbative processes, such as tunneling to other vacua, black hole nucleation and/or collapse of the domain wall network. Therefore, we may expect a finite (but exponentially long) lifetime from nonperturbative instabilities, even in perturbatively stable models. An example of such an instanton was found in \cite{Mithani:2011en}, following the first version of our work. Assuming that the classical theory is valid for arbitrarily small scale factors, they constructed a Euclidean solution where the simple harmonic universe with $\gamma \sim 1$ tunnels to $a \to 0$ with a rate $P \sim \exp(-3/(16 G^2 |\Lambda|))$. In this case, the universe would be metastable, with an exponentially long lifetime.

However, it is important to stress that the instanton of \cite{Mithani:2011en} is singular, and both its existence and the predicted value of the decay rate may depend on physics at some UV (or even the Planck) scale.  Furthermore, it is not clear whether this solution gives the leading contribution to the decay rate.  Although we may expect that there are nonperturbative instabilities, a full analysis will require a concrete (possibly UV-complete) model  for the simple harmonic universe, which would be an interesting direction for future work.

{\it Conclusions and Questions.} 
Our model with $\gamma \sim 1$ seems to provide an example of an eternal universe without singularities.  This universe is both classically and quantum mechanically stable against small perturbations at linearized level.  It avoids many problems with eternal bouncing cosmologies   \cite{extraNECviol, Senatore, starobinsky,brandenberger2, others}.
Possibly, however, the background ``solid" could have microscopic dynamics that produce entropy, leading to a singularity even in our seemingly eternal models.  This is an interesting, but model-dependent, question. We have focused on model-independent bounds here.
This raises the question, can we prove a `quantum singularity theorem' that applies to closed Universes, without assuming unphysical energy conditions? 

The cyclic nature of these cosmologies strongly suggests
searching for exactly periodic quantum states in our geometry.  Could some of these special quantum states be eternal, and provide ``natural" boundary conditions for certain closed cosmologies,
in analogy with \cite{HH}?

Can we embed
realistic $\Lambda$CDM cosmologies, with a preceding phase of inflation, into the expansion phase of one of our cycles in the $\gamma \ll 1$ case?  This would require a transition
from radiation/matter dominance during expansion to curvature/``solid" dominance near the following bounce.  Given their relative scalings with $a$, this may require the radiation and matter modes to be ``Higgsed" above a large energy scale.
As we have seen, such a universe with $\gamma \ll 1$ appears unstable.
However, we were maximally pessimistic in ignoring free streaming; could this effect vitiate the growth of inhomogeneous perturbations? 
Alternatively, for the stable, eternal $\gamma \sim 1$ cosmologies, can we envision a Universe which begins in such a phase, persists there for a long period, and then transitions to a realistic inflationary Universe \cite{Biswas:2011qe}?  Could either of these possibilities demonstrate that our observed universe might not have emerged from an initial singularity \cite{uslong}?

We thank R.\ Bousso,  S. \ Dubovsky, W.\ Fischler, E.\ Flanagan, B.\ Freivogel, G.\ Horowitz, N.\ Kaloper, A.\ Linde, L.\ McAllister,A.\ Moradinezhad, H.\ Murayama, P.\ Shellard, S.\ Shenker, E.\ Silverstein,  L.\ Susskind, G.\ Villadoro, B.\ Xue, and S.\ Yaida for valuable discussions.  This research is supported in part by NSF grants PHY-02-44728, PHY-05051164, by the US DOE under contract DE-AC02-76SF00515, and by ERC grant BSMOXFORD no.
228169.  B.H.\ is also supported by a William K. Bowes Jr.\ Stanford Graduate Fellowship.

\end{document}